\documentclass[aps,prl,twocolumn,superscriptaddress,groupedaddress]{revtex4}
\usepackage{graphicx}
\usepackage{amsmath}
\usepackage{float}
\usepackage{color}

\newcommand {\NF}{{N_{\rm F}}}

\newcommand {\bbk}{{\bf k}}
\newcommand {\bbq}{{\bf q}}
\newcommand {\bbkk}{{\bf {k'}}}

\newcommand {\on}{{\omega_n}}
\newcommand {\onp}{{\omega_{n'}}}
\newcommand {\afo}{{\alpha^{2} F(\omega)}}
\newcommand {\gkkl}{{g_{\bf k \bf {k'}}^{\nu}}}

\newcommand {\deltakk}{{\delta(\epsilon_{\bf {k'}})}}
\DeclareMathOperator{\sech}{sech}

\begin{document}

\title{Electron-phonon coupling and pairing mechanism in $\beta$--Bi$_2$Pd centrosymmetric superconductor}

\author{Jing-Jing Zheng}
\affiliation{Department of Physics, Applied Physics and Astronomy,
Binghamton University-SUNY, Binghamton, New York 13902, USA}
\affiliation{Institute of Theoretical Physics and Department of
Physics, Shanxi University, Taiyuan 030006, People's Republic of
China}
\author{E. R. Margine }
\thanks{Corresponding author. E-mail address: rmargine@binghamton.edu (E.R. Margine).}
\affiliation{Department of Physics, Applied Physics and Astronomy,
Binghamton University-SUNY, Binghamton, New York 13902, USA}

\begin{abstract}
We report first-principles calculations of the superconducting
properties of $\beta$--Bi$_2$Pd within the fully anisotropic
Migdal-Eliashberg formalism. We find a single anisotropic
superconducting gap of $s$--wave symmetry which varies in magnitude
on the different regions of the Fermi surface. The
calculated superconducting energy gap on the Fermi surface, the
superconducting transition temperature, the specific heat, and the quasi-particle density of states are in
good agreement with the corresponding experimental results and
support the view that $\beta$--Bi$_2$Pd is a phonon-mediated single
anisotropic gap superconductor.
\end{abstract}
\maketitle

\section{INTRODUCTION}
In recent years considerable attention has been devoted to the study
of superconductors in the Bi--Pd family of
alloys~\cite{Zhuravlev1,Zhuravlev2,Matthias,Okamoto}, in particular,
the non-centrosymmetric $\alpha$--BiPd~\cite{Jiao,Neupane,Sun,Peets}
and the centrosymmetric
$\beta$--Bi$_2$Pd~\cite{Imai,Herrera,Biswas,Che,Lv,Shein,Sakano,Kacmarcık,Zhao}
compounds that have been found to exhibit topologically protected
surface states with Rashba-like spin splitting. While the
superconducting state of $\alpha$--BiPd appears to be topologically
trivial~\cite{Sun,Neupane}, the structure of the superconducting gap
and the role of the topologically protected surface states in the
superconducting pairing mechanism of $\beta$--Bi$_2$Pd still remain
to be clarified. An early experimental study has suggested that
$\beta$--Bi$_2$Pd is a multigap/multiband superconductor based on
the temperature dependence of the electronic specific heat and the
upper critical magnetic field data~\cite{Imai}. In contrast, recent
scanning tunneling microscopy, calorimetric, Hall-probe
magnetometry, muon spin relaxation, and point-contact spectroscopy
measurements point towards a single $s$--wave superconducting
gap~\cite{Herrera,Kacmarcık,Biswas,Che}. Finally, while the absence
of topological Andreev bound states in point-contact data~\cite{Che}
excludes the possibility of a topological superconducting behavior
at the surface of $\beta$--Bi$_2$Pd, Majorana zero modes have been
identified in $\beta$--Bi$_2$Pd crystalline films~\cite{Lv}.

In this work, we present the first theoretical study of the phonons
and the role of the electron-phonon interaction in the
superconducting state of $\beta$--Bi$_2$Pd by performing
state-of-the-art {\it ab initio} calculations powered by
electron-phonon Wannier
interpolation~\cite{Marzari_RMP,giustino_wannier,Giustino_RMP16}.
Since the strong spin-orbit coupling (SOC) inherent to the heavy
element Bi induces significant changes in the electronic band
structure~\cite{Shein,Sakano}, $\beta$--Bi$_2$Pd compound offers the
opportunity to further study the interplay of the spin-orbit
coupling effect with superconductivity. We find that
$\beta$--Bi$_{2}$Pd displays BCS-like superconductivity with a
single anisotropic superconducting gap in agreement with recent
experimental findings~\cite{Herrera,Biswas,Che,Kacmarcık}.

\section{METHODOLOGY}
The calculations are performed within the generalized gradient
approximation to density-functional theory~\cite{GGA}, employing
fully relativistic norm-conserving pseudopotentials~\cite{nc1,nc2},
as implemented in the {\tt Quantum-ESPRESSO} suite~\cite{QE}.  The
Perdew, Burke, and Ernzerhof (PBE) \cite{PBE} form of the
generalized gradient approximation was chosen to describe the
exchange-correlation energy. The planewaves kinetic energy cutoff is
40~Ry. The Bi $6s^{2}6p^{3}$ and the Pd $4d^9 5s^1$ orbitals were
included as valence electrons. The $\beta-$Bi$_{2}$Pd phase
crystallizes in the centrosymmetric body-centered tetragonal crystal
structure with space group \emph{I}4/\emph{mmm} (No.139). To
facilitate the comparison with the experimental results, we employ
the experimental lattice constants ($a=3.362$ \AA\, and $c=12.983$
\AA~\cite{Zhuravlev1}) for calculations without and with
SOC~\cite{note1}. A similar approach has been used in previous
studies of superconductivity and charge density wave instability of
other layered materials~\cite{Mazin1, Mazin2, Bianco, Chang}. In
both cases, the atomic positions were relaxed using a threshold of
10~meV/\AA\ for the forces. The optimized internal parameter
$z_{Bi}$ for Bi atom is 0.361 (without SOC) and 0.362 (with SOC),
which are in good agreement with the experimentally reported value
of $z_{Bi}=0.363$~\cite{Zhuravlev1}. The electronic charge density
is calculated using an unshifted Brillouin zone (BZ) mesh with
$12^3$ $\bbk$-points  and a Methfessel-Paxton smearing of 0.02~Ry.
The dynamical matrices and the linear variation of the
self-consistent potential are calculated within density-functional
perturbation theory~\cite{baroni2001} on the irreducible set of a
regular $4^3$ $\bbq$-point mesh.

The superconducting gap is evaluated using the anisotropic
Migdal-Eliasberg formalism~\cite{allen_mitrovic,margine_eliashberg}
as implemented in the {\tt EPW}
code~\cite{EPW,margine_eliashberg,Ponce}. We have recently used this
methodology to investigate the superconducting properties of layered
and two-dimensional materials with highly anisotropic Fermi
surfaces~\cite{margine_graphene,margine_Ca-graphene,margine_Li-graphene,Heil}.
The electronic wavefunctions required for the Wannier-Fourier
interpolation~\cite{Marzari_RMP,wannier} in {\tt EPW} are calculated
on a uniform unshifted Brillouin-zone grids of size $8^3$. Eleven
maximally localized Wannier functions, three $p$ states for each Bi
atom and five $d$ states for Pd atom, are included to describe the
electronic structure near the Fermi level. For the anisotropic
Migdal-Eliashberg equations we use $90^3$ $\bbk$-point and $45^3$
$\bbq$-point grids. The Matsubara frequency cutoff is set to 7.5
times the largest phonon frequency, and the Dirac delta functions
are replaced by Lorentzians of widths 25~meV and 0.05~meV for
electrons and phonons, respectively.

The technical details of the Migdal-Eliashberg calculations have
been described extensively in
Refs.~[\onlinecite{margine_eliashberg}] and [\onlinecite{Ponce}],
here we only outline briefly the main procedure. In order to
calculate the superconducting properties the fully anisotropic
Migdal-Eliashberg equations are solved self-consistently along the
imaginary axis at the fermion Matsubara frequencies $\on=(2n+1)\pi
T$ (with $n$ an integer) for each temperature $T$:
\begin{eqnarray} \hspace{-0.25cm}
  &&Z(\bbk,i\on) =
   1 + \frac{\pi T}{\NF \on} \sum_{\bbkk n'}
   \frac{ \onp }{ \sqrt{\onp^2+\Delta^2(\bbkk,i\onp)} } \nonumber \\
  &&\qquad\qquad\qquad\times \deltakk \lambda(\bbk,\bbkk,n\!-\!n'),
  \label{Znorm_surf} \\
 &&Z(\bbk,i\on)\Delta(\bbk,i\on) =
   \frac{\pi T}{\NF} \sum_{\bbkk n'}
   \frac{ \Delta(\bbkk,i\onp) }{ \sqrt{\onp^2+\Delta^2(\bbkk,i\onp)} } \nonumber \\
  &&\qquad\qquad\qquad\times \deltakk \left[ \lambda(\bbk,\bbkk,\!n-\!n')-\mu^*\right].\nonumber\\
\label{Delta_surf}
\end{eqnarray}
$Z(\bbk,i\on)$ is the mass renormalization function,
$\Delta(\bbk,i\on)$ is the superconducting gap function,
$\lambda(\bbk,\bbkk,\!n-\!n')$ is the momentum- and energy-dependent electron-phonon coupling, $\bbk$ ($\bbkk$) is an electronic state of combined band and
momentum index, and $\mu^*$ is the semiempirical Coulomb
parameter~\cite{note3}. The
anisotropic $\lambda(\bbk,\bbkk,\!n-\!n')$ to be used in the
Migdal-Eliashberg equations is given by:
\begin{equation} \label{lambda}
\lambda(\bbk,\bbkk,n - n') = \NF \sum_{\nu} \frac{2\omega_{\bbq
\nu}}{(\on - \onp)^2+\omega_{\bbq \nu}^2} |\gkkl|^2.
\end{equation}
From the superconducting gap function $\Delta(\bbk,i\on)$ one
obtains the superconducting gap at real-valued frequencies
$\Delta(\bbk,\omega)$ via Pad\'{e} approximants~\cite{Pade1,Pade2}.
The superconducting critical temperature $T_c$ is identified as the
temperature at which the leading edge $\Delta(\bbk,\omega=0)$ of the
superconducting gap vanishes.

\begin{figure}[ptb]
\begin{center}
\includegraphics[width=\linewidth]{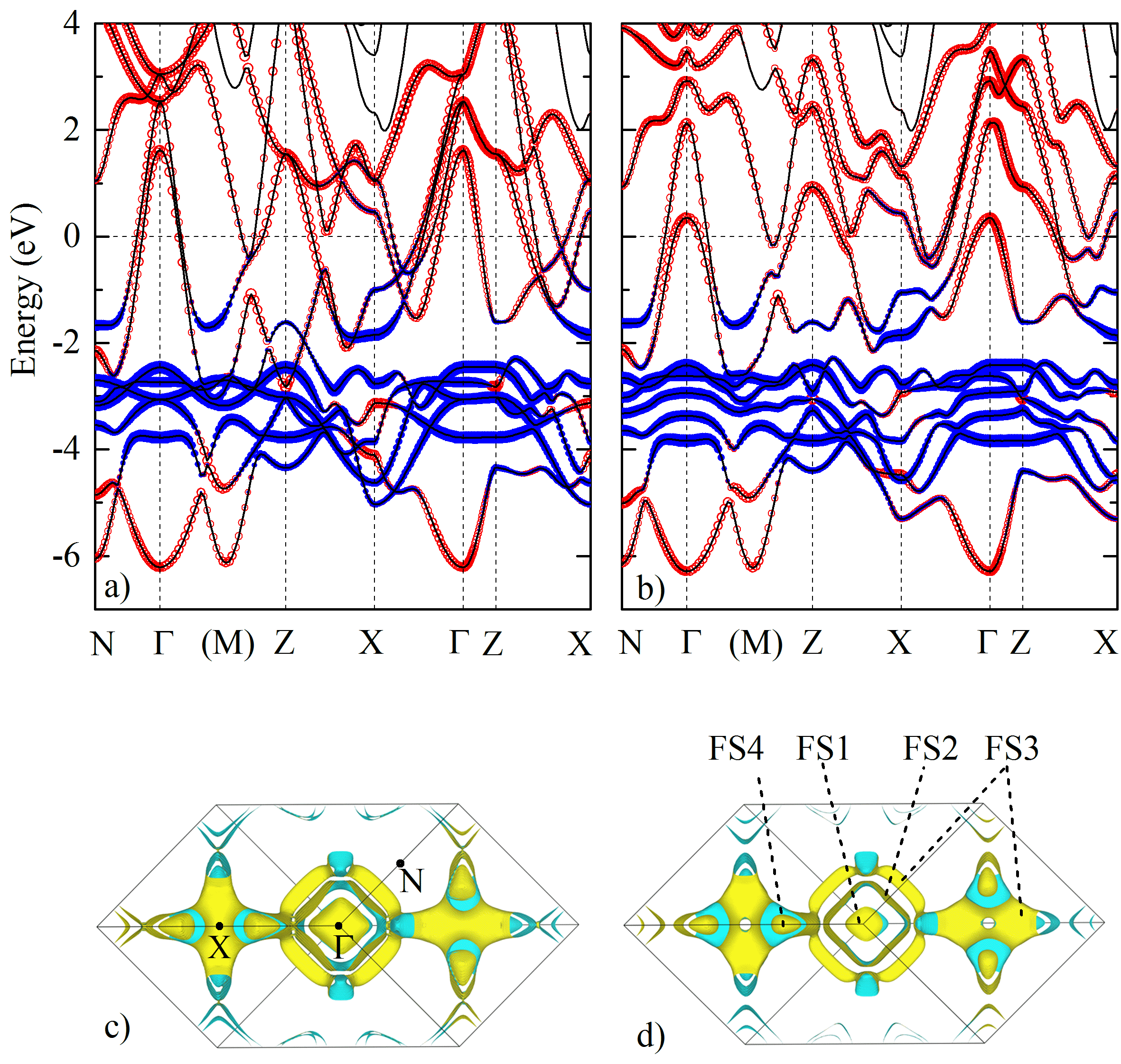}
\end{center}
\caption{ (Color online) Electronic band structure of bulk
$\beta$--Bi$_2$Pd (a) without SOC and (b) with SOC. The size of the
blue and red symbols is proportional to the contribution of Pd-$d$
and Bi-$p$ character. The corresponding Fermi surfaces are plotted
in (c) and (d). The inner surface is shown in cyan, while the outer
part in yellow. The connection between the reciprocal lattice and
the BZ is shown in Supplemental Fig.~S1~\cite{SuppMat}. The hole
pockets are labeled as FS1 and FS2 and the electron pockets are
labeled as FS3 and FS4.}
\label{fig1}%
\end{figure}

\section{ELECTRONIC AND VIBRATIONAL PROPERTIES}
Figures~\ref{fig1}(a) and (b) depict the decomposed band structure
of $\beta$--Bi$_{2}$Pd along the main high-symmetry directions of
the Brillouin zone~\cite{Curtarolo} calculated without and with the
inclusion of spin-orbit coupling. As previously
reported~\cite{Shein,Sakano}, four bands of mixed orbital character
cross the Fermi level. The two hole-like bands centered around the
$\Gamma$ point originate from the Bi--$6p_{x+y}/p_z$ orbitals, while
the two electron-like bands along the $\Gamma$--X direction
originate from the Bi--$6p_{x+y}/p_z$ and Pd--$4d_{xz+yz}/d_{xy}$
orbitals. These bands give rise to a complex multiple sheet Fermi
surface (FS) topology~\cite{Shein} that has been recently resolved
using angular-resolved photoemission spectroscopy~\cite{Sakano}. As shown in Figs.~\ref{fig1}(c)--(d), the
Fermi surface consists of a deformed two-dimensional cylindrical
hole-like sheet (FS2) enclosing a smaller hole pocket (FS1) centered
around the $\Gamma$ point. Surrounding the surface of the outer hole
pocket, there is a three-dimensional electron pocket (FS3) which
confines a second smaller electron pocket (FS4) along the
$\Gamma$--X direction.

The inclusion of the SOC leads to the splitting of bands around the
Fermi level and the opening of two continuous gaps that extend over
the whole Brillouin zone as shown in Fig.~\ref{fig1}(b). As a
result, the electronic density of states at the Fermi level ($N_F$) is
increased by approximately 20\%, from 0.659 to 0.788 states/spin/(eV
unit cell). Finally, there is a noticeable reduction in the size of
the $\Gamma$--centered hole and electron pockets, although the Fermi
surface topology remains unchanged with the inclusion of SOC.

\begin{figure}[ptb]
\begin{center}
\includegraphics[width=0.8\linewidth]{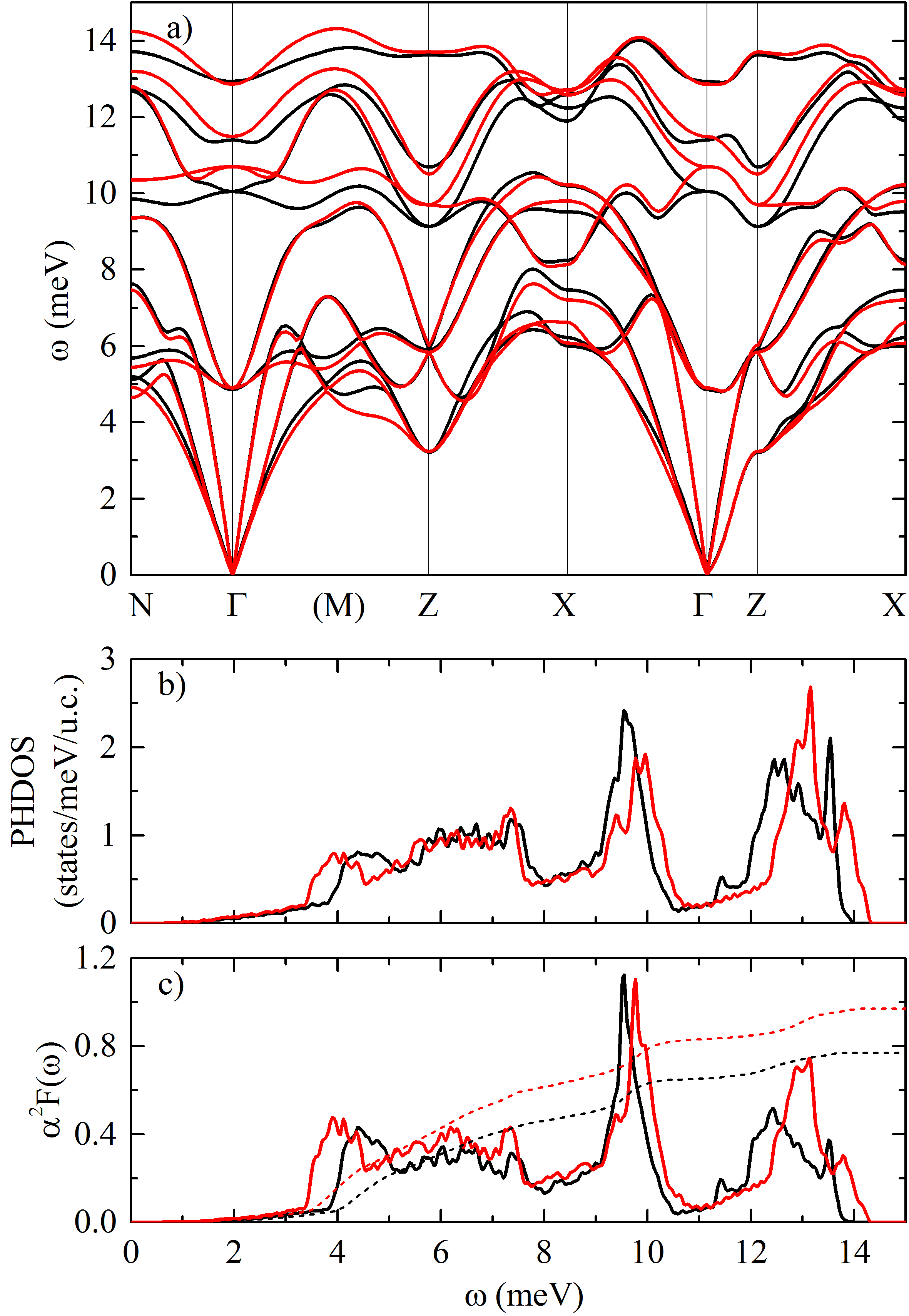}
\end{center}
\caption{(Color online) (a) Phonon dispersion relations and (b)
phonon density of states of $\beta$--Bi$_2$Pd. (c) Isotropic
Eliashberg spectral function $\afo$ (solid line) and cumulative
electron-phonon coupling strength $\lambda(\omega)$ (dashed line). The
data without and with SOC are presented in black and red,
respectively.}%
\label{fig2}%
\end{figure}

In Figs.~\ref{fig2}(a) and (b), we compare the phonon dispersion
relations and the phonon density of states (PHDOS) calculated
without and with SOC. In both cases, three regions can be clearly
distinguished in the PHDOS: a low-energy region that extends up to
8.0~meV, an intermediate region from 8.0 to 10.0~meV, and a
high-energy region above 10.0~meV. The decomposition of the phonon
spectrum with respect to atomic vibrations is provided in
Supplemental Fig.~S2~\cite{SuppMat}. The three acoustic phonon
branches have a mixed character with both types of atoms involved in
the lattice vibration. That also holds for the first two optical
branches. The higher optical phonons stem mostly from the vibration
of Pd atoms, consistent with their lighter mass. The phonon spectrum
only changes slightly with the inclusion of SOC. The frequencies of
the highest optical modes are pushed upward by about 0.36~meV
(2.5~\%) across the whole BZ, while the lowest acoustic branch
softens
 along the $\Gamma$--Z direction.

\begin{figure*}[t]
\begin{center}
\includegraphics[width=0.9\linewidth]{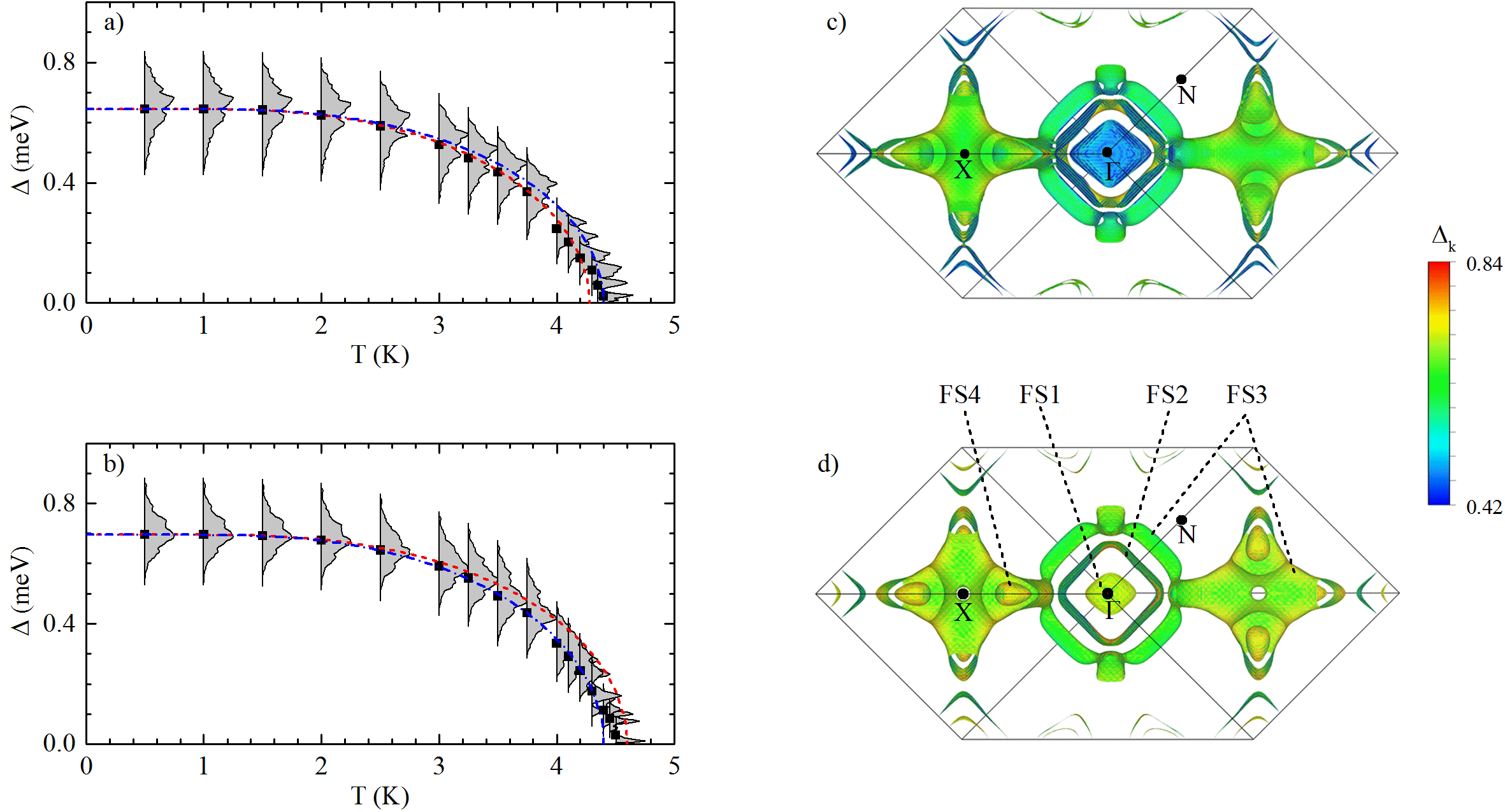}
\end{center}
\caption{(Color online) Energy distribution of the anisotropic
superconducting gap $\Delta_{\mathbf{k}}$ of $\beta$--Bi$_2$Pd as a
function of temperature (a) without SOC and (b) with SOC. The gap is
calculated using a Coulomb pseudopotential $\mu^{*}$ of 0.1. The
black squares represent the average value of the gap which vanishes
at the critical temperature $T_c=4.40$~K (without SOC) and 4.55~K
(with SOC). The red dashed and blue dashed dotted lines are fits to
the calculated data based on the BCS model and $\alpha$ model,
respectively. Momentum-resolved superconducting gap
$\Delta_{\mathbf{k}}$ (in meV) on the Fermi surface of
$\beta$--Bi$_2$Pd at 1~K (c) without SOC and (d) with SOC.}%
\label{fig3}%
\end{figure*}

In order to quantify the interaction between electrons and phonons,
we calculate the Eliashberg spectral function, $\afo$, and the
cumulative electron-phonon coupling (EPC) strength,
$\lambda(\omega)$. As shown in Fig.~\ref{fig2}(c), the Eliashberg
spectral function follows the trend of the PHDOS and exhibits a
similar peak structure. Although the phonons from all energy regions
contribute to the total EPC, the relative contributions of each set
of modes differ considerably. The low-energy modes (up to 8.0~meV)
account for approximately 60\% of the total coupling
($\lambda=0.77$), while the phonons in the intermediate- and
high-energy regions have a less sizable contribution of 22\% and
18\%, respectively. With inclusion of SOC, $\lambda$ increases by
26\% to 0.97. The SOC-induced enhancement of $\lambda$ correlates
with the increased density of states at the Fermi level and the
softening of the low-energy phonon modes. In particular, the
phonons in the low-energy region account for 0.15 of the total
increase in $\lambda$, while the phonons in the intermediate- and
high-energy regions only account for 0.01 and 0.04, respectively.

\section{SUPERCONDUCTING PROPERTIES}
After assessing the EPC, we investigate its effect on the
superconducting pairing mechanism. The superconducting gap function
and the critical temperature are estimated by solving the
anisotropic Migdal-Eliashberg
equations~\cite{allen_mitrovic,margine_eliashberg,Ponce}.
Figs.~\ref{fig3}(a)--(b) show the energy distribution of the
superconducting gap at the Fermi level as a function of temperature
calculated without and with SOC and for a typical Coulomb
pseudopotential $\mu^*=0.1$. We find that $\beta$--Bi$_2$Pd exhibits
an anisotropic single-gap structure characterized by a relatively
broad energy profile with a spread of 0.38~meV (without SOC) and
0.34~meV (with SOC). The average value of the gap at zero
temperature is estimated to be $\Delta(0)=0.65$~meV (without SOC)
and $\Delta(0)=0.70$~meV (with SOC). These values are in good
agreement with scanning tunneling microscopy and muon spin
relaxation measurements yielding 0.76~meV~\cite{Herrera} and
0.78~meV~\cite{Biswas}, respectively. Slightly larger values of
0.88~meV and 0.92~meV were reported based on
calorimetric and Hall-probe magnetometry
measurements~\cite{Kacmarcık,note4}, in the first case, and
point-contact spectroscopy measurements~\cite{Che}, in the second
case. It is noteworthy that scanning tunneling measurements
performed on the same samples as used for the calorimetric and
Hall-probe magnetometry measurements returned a superconducting gap
in the 0.77--0.82~meV range~\cite{Kacmarcık,note5}. These
differences in the experimental results still remain
to be understood, with one possible cause the surface proximity
effect~\cite{Kacmarcık}. Based on the Migdal-Eliashberg results,
the superconducting gap vanishes at a critical temperature
$T_c=4.40$~K (without SOC) and $T_c=4.55$~K (with SOC), close to the
lower limit of the reported experimental values ranging from 4.25~K
to
5.4~K\cite{Zhuravlev1,Zhuravlev2,Imai,Herrera,Biswas,Che,Sakano,Kacmarcık}.
In order to test the sensitivity of our results to the parameters
entering in the Migdal-Eliashberg equations, we have repeated the
calculations for different electron and phonon meshes, smearing
values, and Matsubara frequency cutoffs. The superconducting gap and
critical temperature are found to be converged within 5\%.

Figures~\ref{fig3}~(c) and (d) show the superconducting gap at 1~K
on different parts of the Fermi surface. The same figures under
three different perspectives are shown in Supplemental
Fig.~S3~\cite{SuppMat}. We can clearly see that the gap opens on
both the electron and hole pockets, and the magnitude of the gap
changes greatly on some of the Fermi surface sheets. An
essentially isotropic distribution is observed on the Bi-dominated
inner hole FS1 pocket and the central region of the
three-dimensional electron FS3 sheet with mixed Bi and Pd orbital
character. On the contrary, the other parts of the Fermi surface
display a strong anisotropy, in particular, the nearly cylindrical
FS2 sheet shows the largest spread in $\Delta_{\mathbf{k}}$.
Although the topology of the superconducting gap on the Fermi
surface is practically not modified by the inclusion of the SOC,
there is a noticeable difference between the absolute value of the
gap on the inner Fermi surface. In the latter case, the
superconducting gap is no longer characterized by a small spread
around the minimum value, but around the average value of
$\Delta_{\mathbf{k}}$. This change is likely driven by the softening
of the lowest acoustic branch along the $\Gamma$--Z direction which
involves mainly out-plane Bi$_z$ phonons. The structure of the gap
correlates closely with the variation of the electron-phonon
coupling strength on the individual Fermi surface sheets shown in
Supplemental Fig.~S4~\cite{SuppMat}.

We further fit the temperature dependence of the average value of
the superconducting gap using the single-band BCS $s$-wave model~\cite{BCS}
and the single-band $\alpha$ model~\cite{Padamsee}. In the first case, we obtain the
temperature-dependent gap by solving numerically the BCS gap
equation~\cite{BCS,Johnston}

\begin{equation} \label{gap1}
\int^{\frac{k_B \theta_D}{\Delta(0)}}_{0}
\tanh(\frac{\alpha_{BCS}\tilde{E}}{2t})\frac{d\tilde{\varepsilon}}{\tilde{E}}
= \ln[\frac{2k_B \theta_D}{\Delta(0)}],
\end{equation}
with $\Delta(0)$ taken from our first-principles calculations. Here $\alpha_{BCS} = 1.764$,
$t=T/T_c$ is the reduced temperature, $\tilde{\varepsilon}=
\varepsilon/\Delta(0)$ is the normalized normal-state single-particle energy, $k_B \theta_D$
is the maximum phonon energy within the Debye theory set to 1000,
$\tilde{E}=\sqrt{\tilde{\varepsilon}^{2}+\tilde{\Delta}^{2}}$ is the
normalized excited quasi-particle (electron and hole) energy, and
$\tilde{\Delta}=\Delta/\Delta(0)$ is the superconducting
gap normalized by its zero temperature value. In the second case, the data is fitted using the single-gap
$\alpha$ model~\cite{Padamsee,Johnston} which extends the BCS theory by
introducing an adjustable parameter $\alpha=\Delta(0)/k_B T_c$. In
this model the reduced gap $\tilde{\Delta}$ is assumed to be the same
as in the BCS theory calculated from Eq.~(\ref{gap1}).

The results obtained with the two models are shown in
Figs.~\ref{fig3}(a)--(b) as red dashed and blue dashed dotted lines.
Within the $\alpha$ model, the fitting parameters are found to be
$\alpha = 1.71$  and $\alpha = 1.85$ for the calculations without
and with SOC. All fitting curves give an overall good description of
the temperature dependence of the superconducting gaps, although
noticeable deviations can be seen in the region around the critical
temperature. In particular, the two sets of {\it ab initio} data are
best reproduced by the BCS model for calculations without SOC and by
the $\alpha$ model for calculations with SOC, respectively.

\begin{figure}[ptb]
\begin{center}
\includegraphics[width=0.8\linewidth]{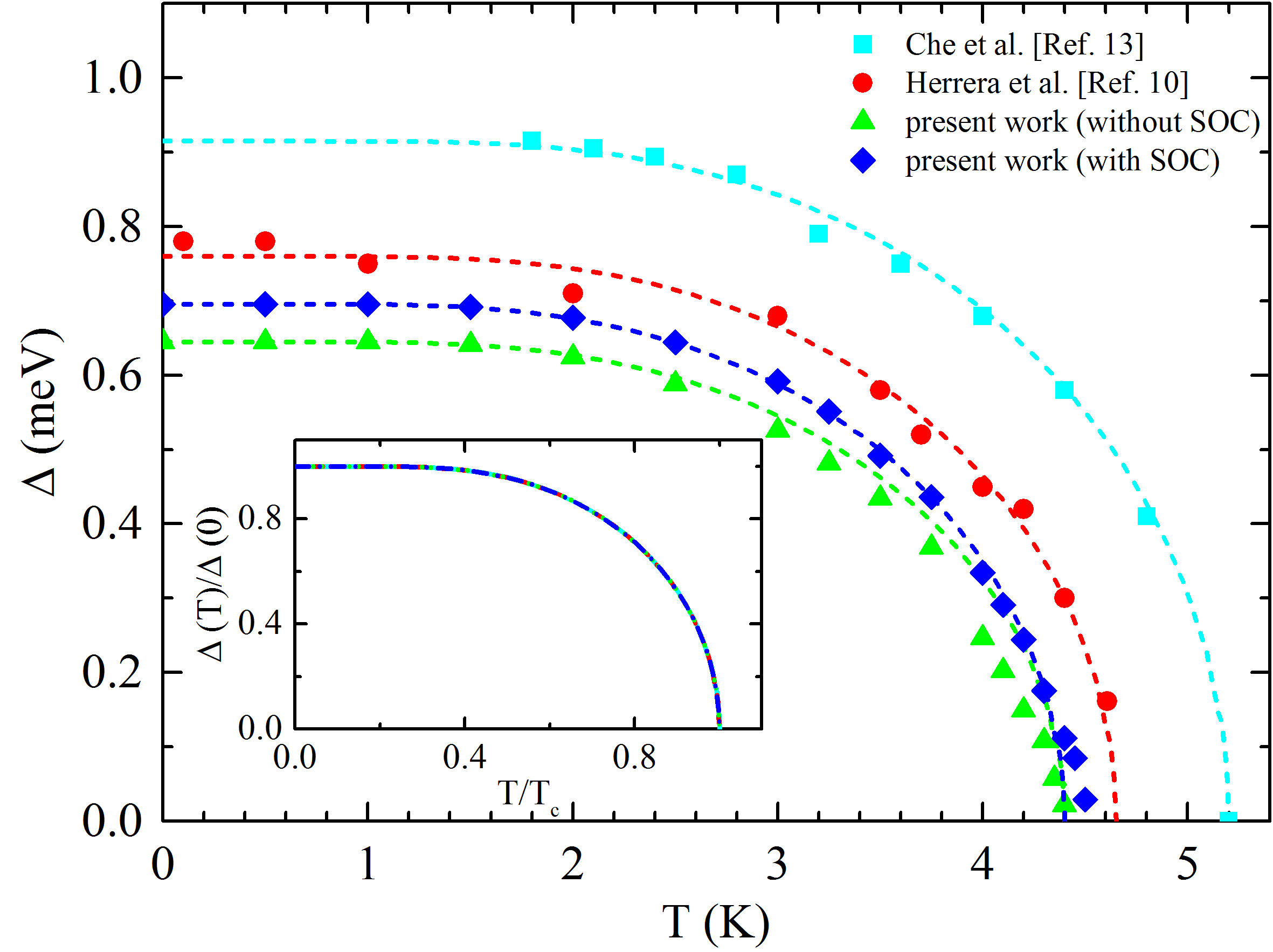}
\end{center}
\caption{ (Color online) Comparison between theoretical and
experimental~\cite{Che,Herrera} superconducting gaps as a function
of temperature. The theoretical data represent the average value of
the gap extracted from
Figs.~\ref{fig3}(a) and (b). The curves are fits to the superconducting gap values within the $\alpha$ model.}%
\label{fig4}%
\end{figure}

In Fig.~\ref{fig4} we compare our results for the superconducting gap with the experimental data
extracted from scanning tunneling microscopy~\cite{Herrera} and
point-contact spectroscopy~\cite{Che} measurements. We find that
both sets of experimental gaps are best fitted with the $\alpha$
model, similarly to our SOC calculations.  We obtain $\alpha = 1.90$
for the tunneling conductance data and $\alpha = 2.05$ for
point-contact data, respectively. For comparison, the fits obtained
with the BCS model display much larger deviations as shown in
Supplemental Fig.~S5~\cite{SuppMat}. The overall
agreement between the theory and experiment is good, the temperature
behavior of the superconducting gap (i.e., the shape of the curve)
is almost perfectly reproduced. This can be more clearly seen in the
inset in Fig.~\ref{fig4}, where we plot the normalized
superconducting gap $\Delta(T)/\Delta(0)$ versus the reduced
temperature $T/T_c$.

\begin{figure}[ptb]
\begin{center}
\includegraphics[width=0.8\linewidth]{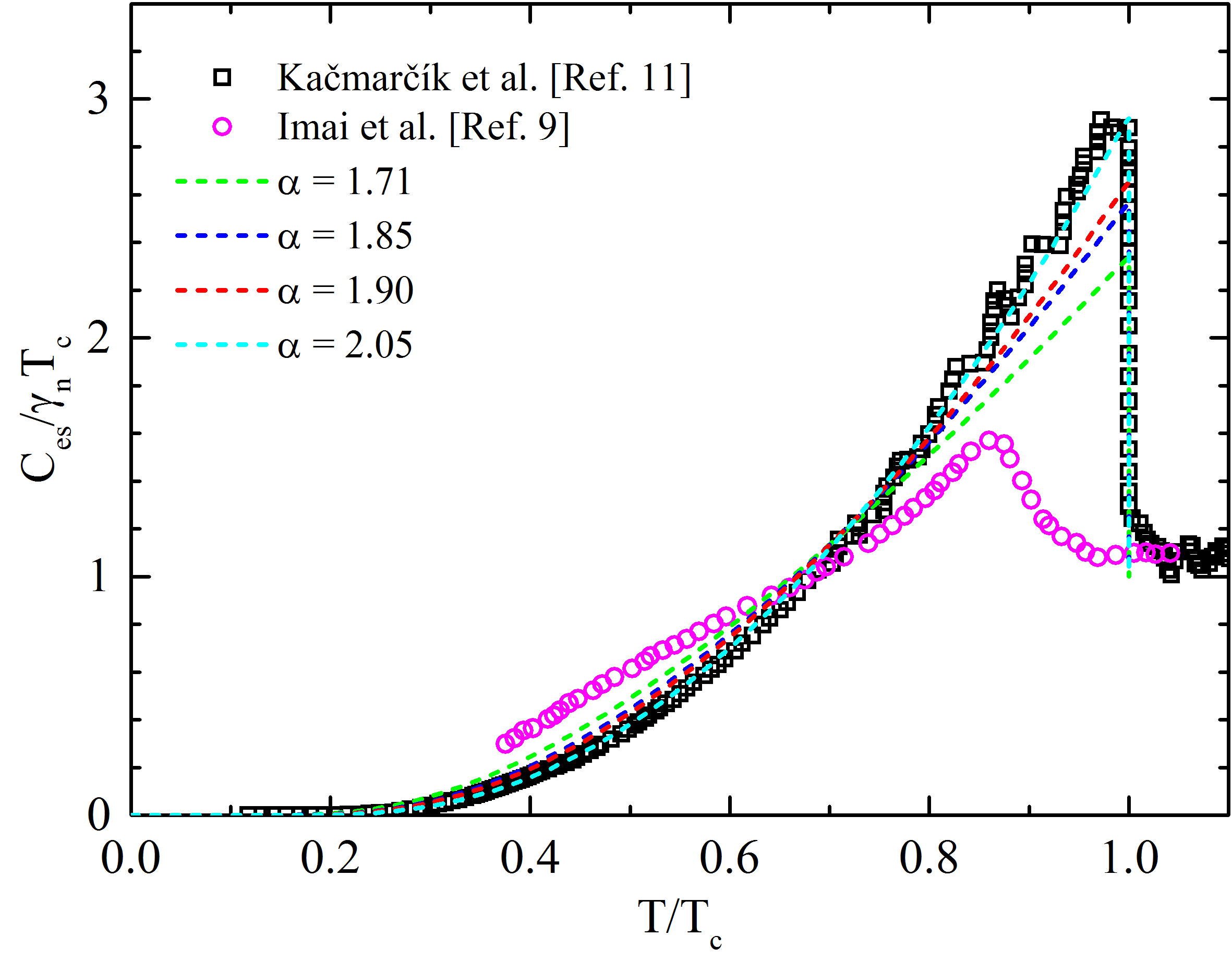}
\end{center}
\caption{ (Color online) Calculated and experimental normalized
electronic specific heat as a function of $T/T_c$.}%
\label{fig5}%
\end{figure}

Next, the temperature dependence of the normalized electronic specific heat in the superconducting state is calculated within the
$\alpha$ model using the following expression~\cite{Johnston}:

\begin{equation} \label{specific heat}
\frac{C_{es}(t)}{\gamma_{n}T_{c}} =
\frac{6\alpha^3}{\pi^{2}t}\int^{\infty}_{0}[f(1-f)(\frac{\tilde{E}^2}{t}-\frac{1}{2}\frac{d\tilde{\Delta}^2}{dt})]d
\tilde{\varepsilon},
\end{equation}
where $f= \left[\exp(\alpha \tilde{E}/t)+1\right]^{-1}$ is the
Fermi-Dirac distribution function and $\gamma_n =(2/3) \pi^2 k_B^2
\NF$ is the Sommerfeld coefficient. The temperature derivative of
the gap function $d\tilde{\Delta}^{2}(t)/dt$ is given
by~\cite{Johnston}:

\begin{equation} \label{specific heat}
\frac{d\tilde{\Delta}^{2}(t)}{dt} =
\frac{\int^{\infty}_{0}\sech^2(g)d\tilde{\varepsilon}}{\int^{\infty}_{0}[\frac{t\sech^2(g)}{2(\tilde{\varepsilon}^2+\tilde{\Delta}^2)}-\frac{t^2\tanh(g)}{\alpha_{BCS}(\tilde{\varepsilon}^2+\tilde{\Delta}^2)^{3/2}}]d\tilde{\varepsilon}},
\end{equation}
where $g =
\frac{\alpha_{BCS}\sqrt{\tilde{\varepsilon}^2+\tilde{\Delta}^2}}{2t}$.
In Fig.~\ref{fig5} we plot the reduced specific heat obtained from
the temperature dependence of the superconducting gap corresponding
to the four fits based on the $\alpha$ model in Fig.~\ref{fig4}
together with the experimental data extracted from
thermal-relaxation~\cite{Imai} and calorimetric~\cite{Kacmarcık}
measurements. To facilitate the comparison, we renormalize the
temperature to $T_c$, using the corresponding experimental and
theoretical values. The normalization procedure for the experimental
data is shown in Supplemental Fig.~S6~\cite{SuppMat}. The results
for $\alpha < 2$ associated with the theoretical data (green and
blue dashed lines) and scanning tunneling microscopy data (red
dashed line) compare well over the whole temperature range. The
discontinuity at the critical temperature $\Delta C_e/\gamma_n
T_c=1.426(\alpha/\alpha_{BCS})^{2}$ is found to be 1.34 (for
$\alpha= 1.71$), 1.57 (for $\alpha=1.85$) and 1.65 (for
$\alpha=1.90$), close to the BCS value of 1.43. A larger discrepancy
is observed, particularly in the high-temperature region $t > 0.8$,
when the results for $\alpha < 2$ are compared to the ones for
$\alpha=2.05$ associated with the point-contact data (cyan dashed
line) and the calorimentric specific heat measurements (black
squares)~\cite{Kacmarcık}. In the latter two cases, the jump at the
critical temperature is estimated to be 1.93. Despite the
aforementioned differences, all these results give a specific heat
temperature dependence consistent with the one-gap BCS model.
Moreover, they differ strikingly from the earlier specific heat
measurements (magenta circles),  disproving the suggestion of
multigap superconductivity~\cite{Imai}.

The argument for a multigap scenario presented in
Ref.~[\onlinecite{Imai}] was based on the appearance of an
additional hump in the specific heat at approximately 3~K as shown
in Supplemental Fig.~S6(a). A similar low-temperature feature
observed in the MgB$_2$ superconductor at approximately 10~K was
caused by a second energy
gap~\cite{Bouquet_PRL01,Yang_PRL01,Wang_PhysC01}. Under this
assumption, a two-gap model with energy gap parameters of 0.54~meV
and 1.29~meV~\cite{note2} was shown to reproduce well the
temperature dependence of the measured specific heat~\cite{Imai}.
However, it is noteworthy that in this case the energy value of the
main gap is quite significantly larger than the latest reported
results~\cite{Herrera,Biswas,Che,Kacmarcık}. Moreover,
Ka\u{c}mar\u{c}\'{i}k {\it et al.}~[\onlinecite{Kacmarcık}] very
detailed study of the specific heat capacity could not find any sign
of a second energy gap. Further comparison of the two sets of data
reveals that the jump in the specific heat near the critical
temperature is much broader in Ref.~[\onlinecite{Imai}] compared to
Ref.~[\onlinecite{Kacmarcık}], pointing towards highly homogeneous
Bi$_2$Pd single crystals in the latter case. It is then likely that
the observed hump belongs to other Bi--Pd superconducting alloys and
the X-ray diffraction pattern indeed shows the presence of a
$\alpha$--Bi$_2$Pd phase~[\onlinecite{Imai}].

Besides the temperature dependence of the specific heat, other
measurements have been used to address the issue of multigap
superconductivity. The temperature dependence of the lower critical
field~\cite{Biswas,Kacmarcık,Herrera} and the magnetic field
dependence of the superconducting gap~\cite{Che} were very well
described by one-gap models providing additional support for a
standard single $s$--wave superconducting gap.

\begin{figure}[ptb]
\begin{center}
\includegraphics[width=0.8\linewidth]{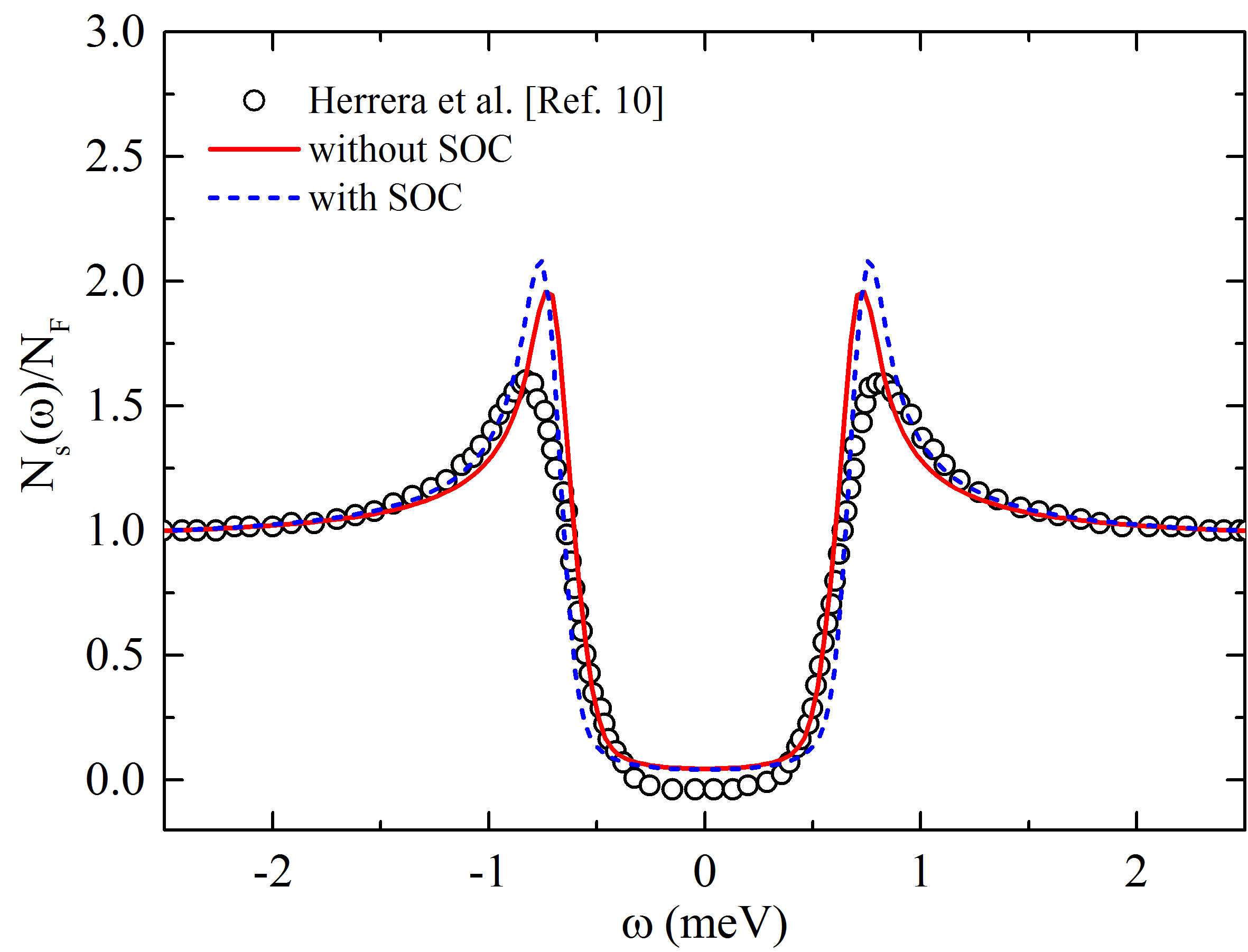}
\end{center}
\caption{ (Color online) Calculated normalized quasi-particle DOS in the
superconducting state of Bi$_2$Pd without (red
line) and with (blue dashed line) SOC compared to tunneling data from Ref.~[\onlinecite{Herrera}] (black circles) for T = 1~K.}
\label{fig6}%
\end{figure}

Finally, we calculate the quasi-particle density of states (DOS) in
the superconducting state $N_s(\omega)$ according to:

\begin{equation} \label{eq.qdos}
\frac{N_S(\omega)}{N_F}= \text{Re} \left[
\frac{\omega}{\sqrt{\omega^2-\Delta^2(\omega)}} \right].
\end{equation}
Our results at 1~K are shown in Fig.~\ref{fig6} and compared
directly to the experimental tunneling conductance of
Ref.~[\onlinecite{Herrera}]. The superconducting DOS is scaled so
that its high energy tail coincides with the DOS in the normal
state. The agreement between our calculations and
experiment is very good, all sets of data exhibiting two symmetric
peaks characteristic of a single-gap structure.

\section{CONCLUSIONS}
In conclusion, we have studied the superconducting properties of
$\beta$--Bi$_2$Pd within the {\it ab initio} anisotropic
Migdal-Eliashberg theory. We have shown that the spin-orbit coupling
only results in a slight increase in the predicted superconducting gap and critical temperature, while the structure of the superconducting gap remains
practically unaffected. We find a sizable anisotropy in the
electron-phonon coupling leading to a single superconducting gap
which varies in magnitude over the disconnected sheets of
the Fermi surface. The experimental superconducting gap, critical
temperature, specific heat, and tunneling conductance are well described assuming a conventional phonon-mediated mechanism as the origin of the
superconducting transition.

\section{ACKNOWLEDGMENTS}
We gratefully acknowledge fruitful discussions with A. Codlea and Z. Pribulov\'{a}. J-J
Zheng acknowledges the support from the China Scholarship Council
(Grant No. 201508140043).


\begin{thebibliography}{99}

\bibitem{Zhuravlev1}
N. N. Zhuravlev, Zh. Eksp. Teor. Fiz. {\bf 32}, 1305 (1957).

\bibitem{Zhuravlev2}
N. N. Zhuravlev, Sov. Phys. Crystallogr. {\bf 3}, 506 (1958).

\bibitem{Matthias}
B. T. Matthias, T. H. Geballe, and V. B. Compton, Rev. Mod.
Phys. {\bf 35}, 1 (1963).

\bibitem{Okamoto}
H. Okamoto, J. Phase Equilib. {\bf 15}, 191 (1994).

\bibitem{Jiao}
L. Jiao, J. L. Zhang, Y. Chen, Z. F. Weng, Y. M. Shao, J. Y. Feng,
X. Lu, B. Joshi, A. Thamizhavel, S. Ramakrishnan, and H. Q. Yuan,
Phys. Rev. B {\bf 89}, 060507(R) (2014).

\bibitem{Neupane}
M. Neupane, N. Alidoust, S.-Y. Xu, I. Belopolski, D. S. Sanchez,
T.-R. Chang, H.-T. Jeng, H. Lin, A. Bansil, D. Kaczorowski, M. Z.
Hasan, and T. Durakiewicz, arXiv:1505.03466v1 (2015).

\bibitem{Sun}
Z. Sun, M. Enayat, A. Maldonado, C. Lithgow, E. Yelland, D. C.
Peets, A. Yaresko, A. P. Schnyder, and P. Wahl, Nat. Commun. {\bf
6}, 6633 (2015).

\bibitem{Peets}
D. C. Peets, A. Maldonado, M. Enayat, Z. Sun, P. Wahl, and A. P.
Schnyder, Phys. Rev. B {\bf 93}, 174504 (2016).

\bibitem{Imai}
Y. Imai, F. Nabeshima, T. Yoshinaka, K. Miyatani, R. Kondo, S.
Komiya, I. Tsukada, and A. Maeda, J. Phys. Soc. Jpn. {\bf 81},
113708 (2012).

\bibitem{Herrera}
E. Herrera, I. Guillam\'{o}n, J. A. Galvis, A. Correa, A. Fente, R.
F. Luccas, F. J. Mompean, M. Garc\'{i}a-Hern\'{a}ndez, S. Vieira, J.
P. Brison, and H. Suderow, Phys. Rev. B {\bf 92}, 054507 (2015).

\bibitem{Kacmarcık}
J. Ka\u{c}mar\u{c}\'{i}k, Z. Pribulov\'{a}, T. Samuely, P.
Szab\'{o}, V. Cambel, J. \u{S}olt\'{y}s, E. Herrera, H. Suderow, A.
Correa-Orellana, D. Prabhakaran, and P. Samuely, Phys. Rev. B {\bf
93}, 144502 (2016).

\bibitem{Biswas}
P. K. Biswas, D. G. Mazzone, R. Sibille, E. Pomjakushina, K. Conder,
H. Luetkens, C. Baines, J. L. Gavilano, M. Kenzelmann, A. Amato, and
E. Morenzoni, Phys. Rev. B {\bf 93}, 220504 (2016).

\bibitem{Che}
L. Che, T. Le, C. Q. Xu, X. Z. Xing, Z. Shi, X. Xu, and X. Lu, Phys.
Rev. B {\bf 94}, 024519 (2016).

\bibitem{Lv}
Y.-F. Lv, W.-L. Wang, Y.-M. Zhang, H. Ding, W. Li, L. Wang, K. He,
C.-L. Song, X.-C. Ma, and Q.-K. Xue, arXiv:1607.07551v1 (2016).

\bibitem{Sakano}
M. Sakano, K. Okawa, M. Kanou, H. Sanjo, T. Okuda, T. Sasagawaa, and
K. Ishizaka, Nat. Commun. {\bf 6}, 8595 (2015).

\bibitem{Shein}
I. R. Shein and A. L. Ivanovskii, J Supercond Nov Magn {\bf 26}, 1
(2013).

\bibitem{Zhao}
K. Zhao, B. Lv, Y. Y. Xue, X. Y. Zhu, L. Z. Deng, Z. Wu, and C. W.
Chu, Phys. Rev. B {\bf 92}, 174404 (2015).

\bibitem{Marzari_RMP}
N. Marzari, A. A. Mostofi, J. R. Yates, I. Souza, and D. Vanderbilt,
Rev. Mod. Phys. {\bf 84}, 1419 (2012).

\bibitem{giustino_wannier}
F. Giustino, M. L. Cohen, and S. G. Louie, Phys. Rev. B {\bf 76},
165108 (2007).

\bibitem{Giustino_RMP16}
F. Giustino, arXiv:1603.06965v1 (2016).

\bibitem{GGA}
J. P. Perdew, Phys. Rev. Lett. {\bf 55}, 1665 (1985).

\bibitem{nc1}
N. Troullier and J. L. Martins,
Phys. Rev. B {\bf 43}, 1993 (1991).

\bibitem{nc2}
M. Fuchs and M. Scheffler,
Comput. Phys. Commun. {\bf 119}, 67 (1999).

\bibitem{QE}
P. Giannozzi {\it et al.}, J. Phys. Condens. Matter {\bf 21}, 395502
(2009).

\bibitem{PBE}
J. P. Perdew, K. Burke, and M. Ernzerhof, Phys. Rev. Lett. {\bf 77},
3865 (1996).

\bibitem{note1}
In agreement with Ref.~[\onlinecite{Shein}], we find that the
optimized lattice parameters with and without the inclusion of SOC
are very closed to each other and less than 2\% larger than the
experimental values. This slight increase in the
lattice constants will only have a minor effect on the magnitude of
the superconducting gap and critical temperature, without affecting
the overall conclusions.

\bibitem{Mazin1}
A. Y. Liu, I. I. Mazin, and J. Kortus, Phys. Rev. Lett. {\bf 87},
087005 (2001).

\bibitem{Mazin2}
M. D. Johannes, I. I. Mazin, and C. A. Howells, Phys. Rev. B {\bf
73}, 205102 (2006).

\bibitem{Bianco}
R. Bianco, M. Calandra, and F. Mauri, Phys. Rev. B {\bf 92}, 094107 (2015).

\bibitem{Chang}
T. Chang {\it et al.}, Phys.
Rev. B {\bf 93}, 245130 (2016).


\bibitem{baroni2001}
S. Baroni, S. de Gironcoli, A. Dal Corso, and P. Giannozzi,
Rev. Mod. Phys. {\bf 73}, 515 (2001).

\bibitem{allen_mitrovic}
P. B. Allen and B. Mitrovi\'{c}, Solid State Phys. {\bf 37}, 1
(1982).

\bibitem{margine_eliashberg}
E. R. Margine and F. Giustino, Phys. Rev. B {\bf 87}, 024505 (2013).

\bibitem{EPW}
J. Noffsinger, F. Giustino, B. D. Malone, C.-H. Park, S. G. Louie,
and M. L. Cohen, Comput. Phys. Commun. {\bf 181}, 2140 (2010).

\bibitem{Ponce}
S. Ponc\'{e}, E. R. Margine, C. Verdi, and F. Giustino,
Comput. Phys. Commun. {\bf 209}, 116 (2016).

\bibitem{margine_graphene}
E. R. Margine and F. Giustino, Phys. Rev. B {\bf 90}, 014518 (2014).

\bibitem{margine_Ca-graphene}
E. R. Margine, H. Lambert, and F. Giustino, Sci. Rep. {\bf 6}, 21414
(2016).

\bibitem{margine_Li-graphene}
J.-J. Zheng and E. R. Margine, Phys. Rev. B {\bf 94}, 064509 (2016).


\bibitem{Heil}
C. Heil, S. Ponc\'{e}, H. Lambert, E. R. Margine, and F. Giustino, Phys. Rev. Lett. (under review).


\bibitem{wannier}
A. A. Mostofi, J. R. Yates, Y.-S. Lee, I. Souza, D. Vanderbilt, and
N. Marzari, Comput. Phys. Comm. {\bf 178}, 685 (2008).

\bibitem{note3}
For a wide range of superconductors, a value of $\mu^*$
in the range 0.1--0.2 has been found, but in several cases values
outside of this range are necessary to explain experimental critical temperatures
\cite{margine_Ca-graphene,Bauer,Subedi_PRB13,Heil}.

\bibitem{Bauer}
J. Bauer, J. E. Han, and O. Gunnarsson, J. Phys.: Condens. Matter
{\bf 24}, 492202 (2012).

\bibitem{Subedi_PRB13}
A. Subedi, L. Ortenzi, and L. Boeri, Phys. Rev. B {\bf 87}, 144504
(2013).

\bibitem{Pade1}
H. J. Vidberg and J. W. Serene, J. Low Temp. Phys. {\bf 29}, 179 (1977).

\bibitem{Pade2}
C. R. Leavens and D. S. Ritchie, Solid State Commun. {\bf 53}, 137
(1985).

\bibitem{Curtarolo}
W. Setyawan, and S. Curtarolo, Comput. Mater. Sci. {\bf 49}, 299
(2010).

\bibitem{SuppMat}
See Supplemental Material at [URL] for Supplemental
Figures S1-S6.

\bibitem{note4}
The superconducting gap was extracted using the data in Ref.~[\onlinecite{Kacmarcık}], namely $2 \Delta/k_B T_c=4.1$ and $T_c=5$~K.

\bibitem{note5}
The superconducting gap was extracted using the data in Ref.~[\onlinecite{Kacmarcık}], namely $2 \Delta/k_B T_c=3.7\pm0.1$ and $T_c=5$~K.


\bibitem{BCS}
J. Bardeen, L. N. Cooper, and J. R. Schrieffer, Phys. Rev. {\bf
108}, 1175 (1957).

\bibitem{Padamsee} H. Padamsee, J. E. Neighbor, and C.
A. Shiffman, J. Low Temp. Phys. {\bf 12}, 387 (1973).

\bibitem{Johnston}
D. C. Johnston, Supercond. Sci. Technol. {\bf 26}, 115011 (2013).

\bibitem{Bouquet_PRL01}
F. Bouquet, R.A. Fisher, N.E. Phillips, D.G. Hinks, and J.D. Jorgensen, Phys. Rev. Lett. {\bf 87}, 047001 (2001).

\bibitem{Yang_PRL01}
H. D. Yang, J.-Y. Lin, H. H. Li, F. H. Hsu, C. J. Liu, S.-C. Li,
R.-C. Yu, and C.-Q. Jin, Phys. Rev. Lett. {\bf 87}, 167003 (2001).

\bibitem{Wang_PhysC01}
Y. Wang, T. Plackowski, and A. Junod, Physica C {\bf 355}, 179 (2001).

\bibitem{note2}
The two energy gap parameters were extracted using the data in Ref.~[\onlinecite{Imai}], namely $2 \Delta_1/k_B T_c=2.5$, $2 \Delta_2/k_B T_c=6$, and $T_c=5$~K.

\end{thebibliography}
\end{document}